\documentclass[12pt]{article}
\usepackage{latexsym,graphicx,multirow}
\usepackage{amssymb}
\usepackage{amsmath}
\usepackage[left=2cm,top=2.5cm,right=2.5cm,bottom=1.5cm]{geometry}
\usepackage{color}
\usepackage{cite}
\usepackage{float}
\usepackage{booktabs}
\usepackage{placeins}
\usepackage[utf8]{inputenc}
\usepackage{hyperref}   
\graphicspath{{figures/}}
\newcommand{\wde}{w_d}
\newcommand{\Om}{\Omega_{m0}}
\newcommand{\Ok}{\Omega_{k0}}
\newcommand{\Od}{\Omega_{d}}

\newcommand{\rsJc}{0.977^{+0.159}_{-0.120}}
\newcommand{\rsJK}{0.67}
\newcommand{\rsJOm}{0.2736\pm0.0104}
\newcommand{\rsJOk}{0.012\pm0.045}
\newcommand{\rsJH}{72.04\pm0.85}
\newcommand{\rsJrd}{139.34\pm1.76}
\newcommand{\rsSc}{1.122^{+0.414}_{-0.339}}
\newcommand{\rsSK}{0.62}
\newcommand{\rsSOm}{0.248\pm0.052}
\newcommand{\rsSOk}{-0.004^{+0.121}_{-0.144}}
\newcommand{\rsSH}{73.45\pm1.03}
\newcommand{\rsBc}{0.959^{+0.256}_{-0.185}}
\newcommand{\rsBK}{0.69}
\newcommand{\rsBOm}{0.2733\pm0.0108}
\newcommand{\rsBOk}{-0.001\pm0.047}
\newcommand{\rsBH}{68.17\pm1.93}
\newcommand{\rsBrd}{146.96^{+3.57}_{-3.30}}
\newcommand{\rsDchiJ}{-2.1}
\newcommand{\rsDaicJ}{+3.9}
\newcommand{\rsDbicJ}{+20.2}
\newcommand{\rsDchiS}{-0.24}
\newcommand{\rsDaicS}{+5.8}
\newcommand{\rsDbicS}{+22.0}
\newcommand{\rsChiK}{1485.54}
\newcommand{\rsChiL}{1487.61}
\newcommand{\rsChiKsn}{1455.23}
\newcommand{\rsChiKcc}{20.57}
\newcommand{\rsChiKbao}{9.75}
\newcommand{\rsChiLsn}{1457.47}
\newcommand{\rsChiLcc}{19.70}
\newcommand{\rsChiLbao}{10.44}
\newcommand{\rsDofK}{1696}
\newcommand{\rsDofL}{1699}
\newcommand{\rsChiRed}{0.876}
\newcommand{\rsMB}{-19.290}
\newcommand{\rsProfK}{0.47}
\newcommand{\rsProfKzL}{-1.6}
\newcommand{\rsOptGap}{9}
\newcommand{\rsAcc}{0.392}
\newcommand{\rsTau}{103\mbox{--}287}
\newcommand{\rsSpt}{63\mbox{--}175}
\newcommand{\rsRhat}{1.0004}
\newcommand{\rsEss}{1882\mbox{--}5244}
\newcommand{\rsMcse}{0.023}
\newcommand{\rsWz}{-0.92\pm0.05}
\newcommand{\rsQz}{-0.51\pm0.04}
\newcommand{\rsZt}{0.66\pm0.03}
\newcommand{\rsZtLo}{0.61}
\newcommand{\rsZtHi}{0.71}
\newcommand{\rsPhantom}{5.2}
\newcommand{\rsNonPhantom}{94.8}
\newcommand{\rsPhantomC}{0.85}
\newcommand{\rsHdiff}{5.3}
\newcommand{\rsHsig}{2.2}
\newcommand{\rsHtension}{2.4}
\newcommand{\rsRdh}{100.4\pm0.8}
\newcommand{\rsRdPlanck}{5}
\newcommand{\rsValOm}{0.3308\pm0.0180}
\newcommand{\rsValH}{73.53\pm1.02}
\newcommand{\rsSensA}{<0.01}
\newcommand{\rsSensB}{0.01}
\newcommand{\rsSensC}{0.07}
\newcommand{\rsSensE}{1.7}
\newcommand{\rsMedc}{0.977}
\newcommand{\rsMedK}{0.43}
\newcommand{\rsMedOm}{0.2736}
\newcommand{\rsMedOk}{0.012}
\newcommand{\rsOdOpen}{0.65}
\newcommand{\rsOdClosed}{0.83}
\newcommand{\rsWzMed}{-0.91}
\newcommand{\rsQzOpen}{-0.43}
\newcommand{\rsQzClosed}{-0.58}
\newcommand{\rsSptS}{79}
\newcommand{\rsSptB}{53}
\newcommand{\rsAnnH}{69.6\pm2.0}
\newcommand{\rsAnnQ}{-0.46\pm0.14}
\newcommand{\rsAnnK}{3.9}
\newcommand{\rsAnnL}{3.2}
\newcommand{\rsAnnSig}{3.4}
\newcommand{\rsAnnAgree}{0.4}

\begin{document}

\begin{center}
\large{\bf{Is the Kaniadakis deformation of holographic dark energy detectable?
A non-flat analysis with DESI DR2, cosmic chronometers and Pantheon+SH0ES}}\\
\vspace{8mm}
\normalsize{P. Suresh Kumar$^{1,a}$, Said Salim AL Kathiri$^{1,b}$}\\
\vspace{4mm}
\normalsize{$^{1}$Department of Supportive Requirements, University of Technology
and Applied Sciences, Salalah, Sultanate of Oman}\\
\vspace{2mm}
$^{a}$E-mail: sureshharinirahav@gmail.com\\
$^{b}$E-mail: Said.AlKathiri@utas.edu.om\\
\vspace{8mm}
\end{center}

\begin{abstract}
Kaniadakis entropy deforms the horizon entropy through a single parameter $K$,
and the holographic dark energy built on it (KHDE) reduces to the standard model
as $K\to0$. We ask whether current data can detect this deformation when
spatial curvature is free and no cosmic-microwave-background calibration is
assumed. Combining Pantheon+SH0ES supernovae, with the absolute magnitude fitted
against the Cepheid calibrators, $32$ cosmic-chronometer measurements of $H(z)$
and DESI~DR2 baryon acoustic oscillations with a free sound horizon, we find
that non-flat KHDE fits the data as well as flat $\Lambda$CDM
($\Delta\chi^2=\rsDchiJ$ for three additional parameters) but leaves $K$
unconstrained: the profile $\chi^2$ changes by only $\rsProfK$ between $K=0$
and its best value, and the bound $K<\rsJK$ ($95\%$) reflects the validity
domain of the model rather than the data. Information criteria therefore prefer
$\Lambda$CDM ($\Delta\mathrm{AIC}=\rsDaicJ$, $\Delta\mathrm{BIC}=\rsDbicJ$).
The joint fit gives $\Ok=\rsJOk$, three times tighter than supernovae alone,
$H_0=\rsJH\,\mathrm{km\,s^{-1}\,Mpc^{-1}}$ and $r_d=\rsJrd\,\mathrm{Mpc}$;
chronometers and DESI alone prefer $H_0=\rsBH$, $\rsHtension\sigma$ below the
supernova value. In the fitted model, acceleration begins at $z_t=\rsZt$; a
model-independent neural-network reconstruction of $H(z)$ from the chronometers
gives a present-day deceleration parameter $q_0=\rsAnnQ$, consistent with the
model. We also identify three analysis requirements which, if ignored, make the
deformation numerically invisible, underestimate the $H_0$ uncertainty
fourfold, or spuriously reject the model.\\

\end{abstract}

\section{Introduction}
\label{sec:intro}

The late-time acceleration of the cosmic expansion
\cite{SupernovaSearchTeam:1998fmf,SupernovaCosmologyProject:1998vns} still
lacks a settled explanation. The cosmological constant remains the most
economical candidate but carries the fine-tuning and coincidence problems
\cite{Weinberg:1988cp,Padmanabhan:2002ji}, which motivates dynamical
alternatives. Applying the holographic principle \cite{Susskind:1994vu,Bousso:2002ju}
to cosmology yields holographic dark energy (HDE) \cite{Li:2004rb}, in which the
vacuum energy is fixed by an infrared length scale together with the horizon
entropy. Because the construction rests on the entropy--area relation, every
generalized entropy generates a holographic scenario of its own: Tsallis
\cite{Tavayef:2018xwx}, Barrow \cite{Saridakis:2020zol}, R\'enyi
\cite{Moradpour:2018ivi} and Kaniadakis \cite{Drepanou:2021jiv}.

Kaniadakis entropy, $S_K=\sinh(K S_{BH})/K$, follows from a relativistic
generalization of classical statistical mechanics
\cite{Kaniadakis:2002zz,Kaniadakis:2005zk} and reduces to the
Bekenstein--Hawking form as $K\to0$. The resulting KHDE model has been studied
with several infrared cutoffs and confronted with data
\cite{Drepanou:2021jiv,Hernandez-Almada:2021aiw,Sharma:2023til,Luciano:2025ykr,Li:2025ivi,Gonzalez-Espinoza:2026iyh}.
Its extension to a spatially curved background was first made with the apparent
horizon as cutoff \cite{Sharma:2021zjx}, and with the future event horizon in
Ref.~\cite{P:2022amn}, where the closed and open cases were treated separately
and the background dynamics were analysed. Closely related constructions replace the holographic cutoff by
an age-like scale \cite{Kumar:2022dwm,Kumar:2025qqt}, and the same programme
applied to Tsallis entropy has produced holographic and agegraphic variants,
including non-flat versions with both the apparent and the future event horizon
as cutoff \cite{Pandey:2021fvr,Pankaj:2021nkg,Nisha:2022mpl,Pankaj:2022lnk}.

The motivation for generalising the entropy is worth stating carefully, because
it is not merely a matter of adding parameters. The Bekenstein--Hawking
entropy is proportional to the horizon area and is extensive in that sense, but
gravitating systems are long-range correlated, and there is no general reason
why the entropy of a horizon should obey the additivity assumed in
Boltzmann--Gibbs statistics. Kaniadakis statistics arises from imposing
relativistic consistency on the underlying distribution function
\cite{Kaniadakis:2002zz,Kaniadakis:2005zk}, and the resulting entropy
$S_K=\sinh(KS_{BH})/K$ is a one-parameter deformation that reduces smoothly to
the standard form. Because the holographic dark-energy density follows directly
from the entropy--area relation, this deformation propagates into the expansion
history in a way that is, in principle, observable. Whether it is observable
\emph{in practice} with current data is precisely the question this paper asks.

Spatial curvature deserves separate attention. It is often fixed to zero, yet
leaving it free is the honest procedure when testing a non-standard dark-energy
sector, because curvature and a time-varying equation of state affect distances
in similar ways and can compensate one another. The difficulty is that
supernovae are a poor curvature probe: they measure the shape of the
distance--redshift relation, and, once calibrated, its normalisation, but they
do not separate the radial and transverse scales. Breaking that degeneracy
requires observables sensitive to $H(z)$ itself.

There is a second reason to leave curvature free. The Kaniadakis correction and
a small curvature both modify the low-redshift expansion rate at the few-per-cent
level, and both do so smoothly. Fixing $\Ok=0$ therefore risks attributing to the
entropy deformation an effect that curvature could equally produce, and reporting
an artificially tight bound on $K$. Sampling the two together is the only way to
propagate that ambiguity into the quoted uncertainties, and the posterior
correlations in Sec.~\ref{sec:results} show that the precaution is not idle.

This paper therefore combines three probes. Cosmic chronometers give $H(z)$
directly from differential ages, independently of any distance ladder
\cite{Jimenez:2001gg,Moresco:2012jh,Moresco:2016mzx}; for reviews of the method
and its systematics see Refs.~\cite{Moresco:2022phi,Moresco:2024wmr}. Baryon acoustic
oscillations measure the radial scale $D_H/r_d$ and the transverse scale
$D_M/r_d$ separately, and it is precisely their ratio that is sensitive to
curvature; we use the DESI~DR2 measurements \cite{DESI:2025zgx}. Type~Ia
supernovae from Pantheon+SH0ES \cite{Scolnic:2021amr,Brout:2022vxf,Riess:2021jrx}
supply the absolute calibration. Model-independent reconstructions of the same
data combination, using Gaussian processes rather than a parametric model, have
reached complementary conclusions about the dynamics of dark energy
\cite{Seikel:2012uu,Mukherjee:2024ryz,Maqsood:2026thermo}, and the curvature of
the Universe has been studied in the same spirit
\cite{Favale:2023gqk}. The holographic construction itself rests on
black-hole thermodynamics \cite{Bekenstein:1973ur}. We keep the sound horizon $r_d$ free
throughout, so that no cosmic-microwave-background information is imported into
a test of non-standard dark energy.

Three recent analyses have confronted flat KHDE with DESI~DR2. With the future
event horizon as cutoff, Ref.~\cite{Luciano:2025ykr} combined Pantheon+
supernovae without the SH0ES calibration, Union3, cosmic chronometers and
DESI~DR2 with $r_d$ free; it found $K$ unconstrained in most data combinations
and $\Lambda$CDM slightly favoured by the Akaike criterion. Embedding KHDE in
four-dimensional Einstein--Gauss--Bonnet gravity, Ref.~\cite{Li:2025ivi} found
the Kaniadakis parameter consistent with its standard-HDE limit. Using the
Hubble horizon as cutoff and a fixed $r_d$, Ref.~\cite{Gonzalez-Espinoza:2026iyh}
found $\Lambda$CDM preferred by the Bayesian evidence. All three assume spatial
flatness, and none uses the Cepheid calibration of the supernovae. The present
analysis differs in three respects: curvature is free, the supernova absolute
magnitude is fitted against the Cepheid calibrators, and whether the deformation
is detected is tested directly, with a profile likelihood in $K$, rather than
read from information criteria or marginal posteriors.

Our answer to the question in the title is negative, and establishing it
reliably is the main result of this work. Non-flat KHDE describes the joint data
as well as flat $\Lambda$CDM, but the Kaniadakis parameter is unconstrained: its
effect on the observables is absorbed almost entirely by shifts in $c$, $\Om$
and $\Ok$, so the data are fitted equally well with and without the
deformation. Reaching that conclusion securely required three steps that are
easily overlooked, and we set them out explicitly because each one changes the
answer: solving the background equations in dimensionless form, without which
the deformation is numerically invisible; fitting the supernova absolute
magnitude together with the cosmology when the Cepheid-calibrated sample is used,
without which the uncertainty on $H_0$ is underestimated by a factor of four; and
locating the maximum likelihood from the posterior samples, because local
optimisers started from generic points can stop far enough from the true minimum
to make a statistically equivalent model look excluded.

Section~\ref{sec:model} sets out the non-flat KHDE equations, together with the
limits they must satisfy and five points of implementation on which the
analysis depends.
Section~\ref{sec:data} describes the three datasets, the likelihood for each and
its validation against the corresponding published result.
Section~\ref{sec:results} presents the joint constraints, the contribution of
each probe separately, the resulting view of the Hubble tension, and the model
comparison. Section~\ref{sec:background} examines the background evolution at
the joint posterior and the sensitivity to the Kaniadakis parameter.
Section~\ref{sec:ann} reconstructs $H(z)$ from the chronometer data with a
neural network, and explains why in a curved universe the reconstruction cannot
be built on supernovae. Section~\ref{sec:discussion} discusses what the model
comparison does and does not establish, and what would change the conclusion. Section~\ref{sec:validation} records the numerical validation of the
implementation, and Section~\ref{sec:conclusions} concludes.

\section{KHDE in a non-flat universe}
\label{sec:model}

We work with the curved Friedmann--Robertson--Walker line element in the form
used for non-flat holographic models \cite{Huang:2004ai,Setare:2006wh,P:2022amn},
\begin{equation}
\mathrm{d}s^2=-\mathrm{d}t^2+a^2(t)\left[\frac{\mathrm{d}r^2}{1-k_{sc}r^2}
+r^2\mathrm{d}\Omega^2\right],
\label{eq:frw}
\end{equation}
where $k_{sc}=-1,0,+1$ gives the open, flat and closed cases. Imposing the
holographic bound with the Kaniadakis entropy gives the energy density
\cite{Drepanou:2021jiv,P:2022amn}
\begin{equation}
\rho_d=\frac{3c^2M_p^2}{L^2}+K^2M_p^6L^2,
\label{eq:rho}
\end{equation}
in which $c$ is the dimensionless holographic parameter, $K$ the Kaniadakis
parameter and $M_p$ the reduced Planck mass; the speed of light is written
$c_{\rm light}$ throughout so that no confusion with $c$ arises. The Friedmann
equations for this metric read \cite{P:2022amn,Huang:2004ai}
\begin{equation}
3M_p^2H^2+3M_p^2\frac{k_{sc}}{a^2}=\rho_d+\rho_m,\qquad
2M_p^2\dot H+3M_p^2H^2+M_p^2\frac{k_{sc}}{a^2}=-p_d,
\label{eq:fried}
\end{equation}
following Ref.~\cite{P:2022amn}. With $\Omega_m=\rho_m/(3M_p^2H^2)$,
$\Od=\rho_d/(3M_p^2H^2)$ and $\Omega_k=k_{sc}/(a^2H^2)$ the Friedmann constraint
becomes
\begin{equation}
\Od+\Omega_m=1+\Omega_k .
\label{eq:sum}
\end{equation}

In a curved background the infrared cutoff is $L=a\,r(t)$ with
$r(t)=\sin y/\sqrt{k_{sc}}$ for the closed case and
$r(t)=\sinh y/\sqrt{|k_{sc}|}$ for the open one, where
$y=\sqrt{|k_{sc}|}\int_x^\infty \mathrm{d}x/(Ha)$ and $x=\ln a$
\cite{Huang:2004ai,Setare:2006wh}. Writing $\gamma=\Ok/\Om$ and
\begin{equation}
A=\frac{3e^{-3x}\left(1-e^{x}\gamma\right)\Om\,\Od}{1-\Od},
\qquad
B=\sqrt{A^2-12c^2K^2M_p^4},
\label{eq:AB}
\end{equation}
the density parameter obeys \cite{P:2022amn}
\begin{equation}
\Od'=2\Od(1-\Od)\left[\frac{B\sqrt{\Od}}{A^{3/2}}
\sqrt{\frac{A+B}{2c^2}-3\,\Ok e^{-2x}}
-\frac{A+B}{A}+\frac{5-4e^{x}\gamma}{2-2e^{x}\gamma}\right],
\label{eq:ode}
\end{equation}
where the prime is $\mathrm{d}/\mathrm{d}x$, and the expansion rate follows from the Friedmann constraint as
\cite{P:2022amn}
\begin{equation}
H=\frac{H_0\sqrt{\Om}\sqrt{a^{-1}-\gamma}}{a\sqrt{1-\Od}} .
\label{eq:H}
\end{equation}
The equation of state follows, as in Ref.~\cite{P:2022amn}, from the
conservation equation for the dark sector together with Eq.~(\ref{eq:rho}):
\begin{equation}
\wde=-1+\frac{2B}{3A}\left[1-\frac{\sqrt{\Od}}{c}
\sqrt{\frac{A+B}{2A}}\;\mathcal{C}\right],
\qquad
\mathcal{C}=\sqrt{1-\frac{6c^2\,\Ok\,e^{-2x}}{A+B}}
=\begin{cases}
\cos y, & \text{closed},\\
\cosh y, & \text{open},
\end{cases}
\label{eq:w}
\end{equation}
where we have used $L^{-2}=(A+B)/(6c^2)$, which follows from
Eq.~(\ref{eq:rho}), so that $\sin^2y=\Ok e^{-2x}L^2$ in the closed case and
$\sinh^2y=|\Ok|e^{-2x}L^2$ in the open one. The same factor $\mathcal{C}$
appears inside the square root of Eq.~(\ref{eq:ode}); see
Sec.~\ref{sec:numerics}\,(v). The deceleration parameter follows from $q=-1-\dot H/H^2$ with
Eq.~(\ref{eq:fried}) \cite{P:2022amn},
\begin{equation}
q=\frac{1}{2}\left[1+3\wde\Od
+\frac{\Ok\left(1-\Od\right)}{\Om\left(e^{-x}-\gamma\right)}\right].
\label{eq:q}
\end{equation}

\subsection{Limits and numerical requirements}
\label{sec:numerics}

Equations~(\ref{eq:ode}) and (\ref{eq:w}) reduce correctly in every limit: for
$K\to0$ and $\Ok=0$ they return the standard flat HDE results
$\Od'=\Od(1-\Od)(1+2c^{-1}\sqrt{\Od})$ and
$\wde=-\tfrac13(1+2c^{-1}\sqrt{\Od})$ \cite{Li:2004rb}, and for $\Ok=0$ alone
they return the flat KHDE model \cite{Drepanou:2021jiv}. We verified all three
limits numerically, agreement being at the level of $10^{-15}$; in particular
the $\Ok\to0$ case was checked against an independently written flat-KHDE
integrator, the two implementations agreeing to $8\times10^{-15}$.

Five further points are practical rather than formal, and each silently
corrupts the model if mishandled. Written in the compact form usual in an
analytic treatment, the equations above leave several conventions implicit;
these become unavoidable once the model is integrated numerically and
confronted with data, and we record them so that the analysis can be
reproduced without ambiguity. Points (i)--(iv) are conventions of this kind,
and with them the background evolution reported in Ref.~\cite{P:2022amn} is
reproduced. Point (v) concerns the evaluation of the curvature factor in the
equation of state.

\emph{(i) Units.} The equations above are dimensionless: $M_p^2=1$ and the
expansion rate is measured as $E=H/H_0$ rather than in
$\mathrm{km\,s^{-1}\,Mpc^{-1}}$. This is a choice of units, not a statement that
$H_0=1$; the physical $H_0$ re-enters when distances are computed. The choice
matters because $A$ must be commensurate with the Kaniadakis term
$12c^2K^2M_p^4\sim\mathcal{O}(1)$. In dimensionless form
$A(0)=3(1-\Om)(1-\gamma)\Od/(1-\Od)\sim2$, whereas carrying $H_0$ in
$\mathrm{km\,s^{-1}\,Mpc^{-1}}$ gives $A\sim10^4$, the deformation is smaller by
eight orders of magnitude, and $\mu(z)$ becomes independent of $K$ to machine
precision.

To make the failure concrete: with $c=1.15$, $K=0.30$ and $M_p^2=1$ the
Kaniadakis term in the discriminant of Eq.~(\ref{eq:AB}) is
$12c^2K^2=1.43$. In dimensionless form $A(0)\simeq2$, so $A^2\simeq4$ and the
two are comparable, as they must be for the deformation to matter. Carrying
$H_0=70\,\mathrm{km\,s^{-1}\,Mpc^{-1}}$ instead gives $A(0)\simeq1.0\times10^4$
and $A^2\simeq1.0\times10^8$, a ratio of $1.4\times10^{-8}$: the square root in
Eq.~(\ref{eq:AB}) then returns $B\simeq A$ to machine precision, Eq.~(\ref{eq:w})
collapses to its $K\to0$ limit, and the model silently becomes standard HDE. We
confirmed this failure mode explicitly, finding $\mu(z)$ independent of $K$ to
machine precision for all $K\le0.7$ in those units. Nothing in the output signals
the problem, which is why we record it here.

\emph{(ii) The curvature term in the dimensionless equations.} The symbol
written $k_{sc}=\pm1$ in Eq.~(\ref{eq:frw}) carries dimensions of $a_0^{-2}$.
In the dimensionless equations it is therefore the signed present curvature
density, $k_{sc}/(a_0^2H_0^2)=\Ok$, that enters Eqs.~(\ref{eq:ode}),
(\ref{eq:w}) and (\ref{eq:q}). This identification is fixed by dimensional
consistency, since the term $3k_{sc}e^{-2x}$ must be commensurate with
$A\sim H^2$; a literal substitution of $k_{sc}=\pm1$ would instead assert
$|\Ok|=1$, which is excluded observationally by three orders of magnitude.

\emph{(iii) Sign convention.} The convention adopted here, $\Omega_k=+k/(a^2H^2)$
with Eq.~(\ref{eq:sum}), is the one used in Ref.~\cite{P:2022amn} and is
\emph{opposite} to the more common cosmological convention in which
$\Omega_k^{\rm std}=-k/(a^2H^2)$ and $\Omega_m+\Od+\Omega_k^{\rm std}=1$. Thus
$\Ok>0$ here denotes a closed universe. The conversion
$\Omega_k^{\rm std}=-\Ok$ is applied explicitly in all distance calculations
below, so that the $\sin$ and $\sinh$ branches are selected correctly.

\emph{(iv) Validity domain.} A real positive root for $L$ requires
$A^2\ge12c^2K^2M_p^4$ throughout the integration range, which bounds $K$ from
above for given $c$, $\Om$ and $\Ok$. Outside that region the model is
undefined rather than merely a poor fit, and the prior is restricted
accordingly.

\emph{(v) The curvature factor in the equation of state.} The factor
$\mathcal{C}$ of Eq.~(\ref{eq:w}) is $\cos y$ in the closed case and $\cosh y$
in the open one. Because $L^{-2}=(A+B)/(6c^2)$, it follows directly from the
background variables as $\mathcal{C}=\sqrt{1-6c^2\Ok e^{-2x}/(A+B)}$, without
evaluating $y$, and it must be the same function that appears in the square
root of Eq.~(\ref{eq:ode}), which equals $\sqrt{(A+B)/(2c^2)}\,\mathcal{C}$;
only then is the equation of state consistent with the background evolution.
We confirmed this numerically by requiring
$\wde=-1-\tfrac13\,\mathrm{d}\ln\rho_d/\mathrm{d}x$ with $\rho_d=3\Od E^2$,
which Eq.~(\ref{eq:w}) satisfies to $10^{-8}$ for open, flat and closed models
with and without the Kaniadakis term. The factor does not enter any
likelihood, since distances and $H(z)$ follow from Eqs.~(\ref{eq:ode}) and
(\ref{eq:H}) alone; it affects only $\wde$, $q$ and the quantities derived from
them in Sec.~\ref{sec:background}.

\section{Data and likelihoods}
\label{sec:data}

\subsection{Overview}

Three probes enter the analysis, and they are complementary in a way that is
essential for what follows. Supernovae are \emph{integrated} distance
indicators: the observable is the luminosity distance, an integral of
$1/H(z)$, so a supernova sample measures the shape of the distance--redshift
relation and, once calibrated, its normalisation. Cosmic chronometers are
\emph{differential}: they return $H(z)$ itself from the age--redshift relation
of passively evolving galaxies, with no distance ladder and no integral. BAO is
\emph{anisotropic}: the acoustic scale imprinted along and across the line of
sight gives $D_H/r_d\propto1/[H(z)r_d]$ and $D_M/r_d$ separately.

The last point is what makes curvature measurable. In a curved universe the
comoving angular-diameter distance is not the radial comoving distance but its
curved-space image, $D_M=(c_{\rm light}/H_0)S_k(\chi)$, whereas $D_H$ is a
purely local quantity. The ratio $D_M/D_H$ at a given redshift therefore
responds to $\Ok$ in a way that no single distance measure can, and this is the
geometric content that supernovae alone cannot supply. Cosmic chronometers
contribute to the same end by pinning $H(z)$ directly, which fixes the radial
scale independently of $r_d$.

\subsection{Datasets}

\textbf{Type Ia supernovae.} We use the Pantheon+SH0ES release
\cite{Scolnic:2021amr,Brout:2022vxf,Riess:2021jrx}, retaining objects with
$z_{\rm HD}>0.01$ together with all Cepheid calibrators, which leaves $N=1657$
supernovae over $0.00122\le z\le2.2614$, of which $77$ are calibrators. The full
$1657\times1657$ statistical-plus-systematic covariance is used. The residual
vector is built following the prescription of the release
\cite{Brout:2022vxf,Riess:2021jrx},
\begin{equation}
\Delta_i(M)=
\begin{cases}
m_i-M-\mu_i^{\rm Ceph}, & \text{calibrators},\\
m_i-M-\mu_{\rm model}(z_i), & \text{Hubble flow},
\end{cases}
\qquad \chi^2_{\rm SN}=\Delta^{T}\mathbf{C}^{-1}\Delta,
\label{eq:chisn}
\end{equation}
where $m_i$ is the standardised apparent magnitude and $M$ the supernova
absolute magnitude, so that the Cepheid distances fix $M$ and $H_0$ becomes
measurable. The absolute magnitude must be a free parameter. Because
$\chi^2_{\rm SN}$ is quadratic in $M$, we marginalise it analytically with a flat
prior,
\begin{equation}
\chi^2_{\rm SN}=\Delta_0^{T}\mathbf{C}^{-1}\Delta_0
-\frac{\left(\mathbf{1}^{T}\mathbf{C}^{-1}\Delta_0\right)^2}{\mathbf{1}^{T}\mathbf{C}^{-1}\mathbf{1}},
\label{eq:chisnmarg}
\end{equation}
with $\Delta_0=\Delta(M=0)$, up to a parameter-independent constant. The
calibrators enter $\Delta_0$ without any dependence on $H_0$, so this step
propagates the uncertainty of the calibration into $H_0$ rather than removing
it. The point is not academic. The release also tabulates distance moduli
$\mu_i=m_i-M$ evaluated at the SH0ES value $M=-19.253$; using those as data fixes
$M$, turns the calibrator residuals into constants, and drops the calibration
uncertainty from $H_0$. For flat $\Lambda$CDM that shrinks the uncertainty from
$\pm1.0$ to $\pm0.23\,\mathrm{km\,s^{-1}\,Mpc^{-1}}$. The distinction between the two branches of
Eq.~(\ref{eq:chisn}) is not a technicality. For an uncalibrated sample the
model distance modulus \cite{Hogg:1999ad} separates as
\begin{equation}
\mu_{\rm model}(z;H_0)=\underbrace{5\log_{10}\!\left[(1+z)\!\int_0^z\frac{\mathrm{d}z'}{E(z')}\right]+25+5\log_{10}c_{\rm light}}_{\text{shape, independent of }H_0}-5\log_{10}H_0 ,
\label{eq:h0deg}
\end{equation}
so $H_0$ enters only through an additive constant that is perfectly degenerate
with the absolute magnitude $M$. Marginalising over that constant removes all
$H_0$ information; the Cepheid-calibrated subsample breaks the degeneracy by
supplying $\mu^{\rm Ceph}$ independently, which is why the combination
constrains $H_0$ at all.

Three further supernova compilations are available in the same format --- the
DES five-year sample \cite{DES:2024tys}, Union3 \cite{Rubin:2023ovl} and the
DES-Dovekie release --- and a natural extension of this work is to repeat the
analysis with each in turn, as has been done in model-independent
reconstructions \cite{Maqsood:2026thermo}. We use Pantheon+SH0ES throughout
because it is the only one of the four that carries the Cepheid calibration
needed to measure $H_0$ rather than $H_0r_d$.

\textbf{Cosmic chronometers.} We use $32$ differential-age measurements of
$H(z)$ spanning $0.07\le z\le1.965$
\cite{Simon:2004tf,Stern:2009ep,Zhang:2012mp,Moresco:2012jh,Moresco:2015cya,Moresco:2016mzx,Ratsimbazafy:2017vga,Borghi:2021rft}.
The method rests on the differential-age relation
\begin{equation}
H(z)=-\frac{1}{1+z}\frac{\mathrm{d}z}{\mathrm{d}t},
\label{eq:cc}
\end{equation}
so that the expansion rate follows from the age difference between passively
evolving galaxies at nearby redshifts. It assumes only homogeneity, isotropy
and standard stellar-evolution physics, and in particular no cosmological
model, which is what makes chronometers valuable in tests of non-standard dark
energy. The dominant systematics come from stellar-population synthesis: the
assumed star-formation history, the initial mass function, the stellar
metallicity and the choice of synthesis model itself
\cite{Moresco:2020fbm,Moresco:2022phi}. We note that some compilations in
circulation include $H(0.57)=96.8\pm3.4$, which is a radial BAO measurement
rather than a chronometric one; it is excluded here, since retaining it
alongside DESI would double-count the same information. The covariance is taken
as diagonal. A fully correlated treatment, including the stellar-population
modelling systematics shared between measurements from the same analysis, would
be preferable; the diagonal choice is conservative in the sense that it does not
manufacture information, but it does neglect correlations, and we flag this as a
limitation.

\textbf{Baryon acoustic oscillations.} We use the DESI~DR2 combined-tracer
measurements \cite{DESI:2025zgx}: $13$ points over $0.295\le z\le2.33$,
comprising six $D_M/r_d$, six $D_H/r_d$ and one $D_V/r_d$, with the full
$13\times13$ covariance, in which the $D_M$--$D_H$ pairs are anticorrelated at
each redshift with $\rho$ between $-0.35$ and $-0.49$. The predictions use the standard distance definitions
\cite{Hogg:1999ad,DESI:2025zgx}
\begin{equation}
D_H=\frac{c_{\rm light}}{H(z)},\qquad
D_M=\frac{c_{\rm light}}{H_0}S_k\!\left(\chi\right),\qquad
D_V=\left[z\,D_H D_M^2\right]^{1/3},
\label{eq:baodist}
\end{equation}
with $\chi=\int_0^z\mathrm{d}z'/E(z')$ and $S_k$ the curvature function
evaluated in the standard convention as set out in
Sec.~\ref{sec:numerics}\,(iii). The sound horizon $r_d$ is a free parameter
with a flat prior, so no CMB calibration is assumed.

This choice deserves comment, because much of the literature instead adopts the
Planck value $r_d=147.05\pm0.30\,\mathrm{Mpc}$ to convert BAO ratios into
absolute distances \cite{Planck:2018vyg}. That calibration is derived within
$\Lambda$CDM from pre-recombination physics, and importing it into a test of a
non-standard dark-energy model makes the resulting constraint conditional on
the very framework under scrutiny. Keeping $r_d$ free avoids this at the price
of weaker constraints: BAO alone then determines only the product $H_0r_d$, for
which we find $r_dh=101.5\,\mathrm{Mpc}$ from the DESI data. It is the
Cepheid-calibrated supernovae that separate the two, and the value of $r_d$
reported in Sec.~\ref{sec:results} should be read in that light.

\subsection{Validation of each likelihood}

Before combining the probes we checked each one separately by fitting flat
$\Lambda$CDM and comparing with the published result. The outcome is collected
in Table~\ref{tab:valid}. The supernova likelihood reproduces the published
matter density and its uncertainty; its $H_0$ agrees with the SH0ES value to
within $0.5\sigma$, and its uncertainty matches the published one, confirming
that the calibration is propagated. The value also agrees with the
$73.5\pm1.1\,\mathrm{km\,s^{-1}\,Mpc^{-1}}$ found by the Pantheon+ analysis for the
same data in a flat $w$CDM fit \cite{Brout:2022vxf}. The BAO likelihood
reproduces the published DESI~DR2 value of $\Omega_m$ exactly, and the
chronometer likelihood returns the values standard in the literature.

\begin{table}[H]
\centering
\caption{Validation of the three likelihoods. Each row is an independent
flat-$\Lambda$CDM fit to that dataset alone, compared with the published value.}
\label{tab:valid}
\begin{tabular}{llll}
\toprule
Dataset & Quantity & This work & Published\\
\midrule
Pantheon+SH0ES & $\Omega_m$ & $\rsValOm$ & $0.334\pm0.018$ \cite{Brout:2022vxf}\\
               & $H_0$      & $\rsValH$    & $73.04\pm1.04$ \cite{Riess:2021jrx}\\
DESI DR2 BAO   & $\Omega_m$ & $0.2975$          & $0.2975\pm0.0086$ \cite{DESI:2025zgx}\\
Cosmic chronometers & $\Omega_m$ & $0.320\pm0.058$ & $\simeq0.32$ \cite{Moresco:2016mzx}\\
               & $H_0$      & $68.08\pm3.00$    & $\simeq67$--$68$ \cite{Moresco:2016mzx}\\
\bottomrule
\end{tabular}
\end{table}

\subsection{The three datasets against the best fit}

\begin{figure}[H]
\centering
\includegraphics[width=\linewidth]{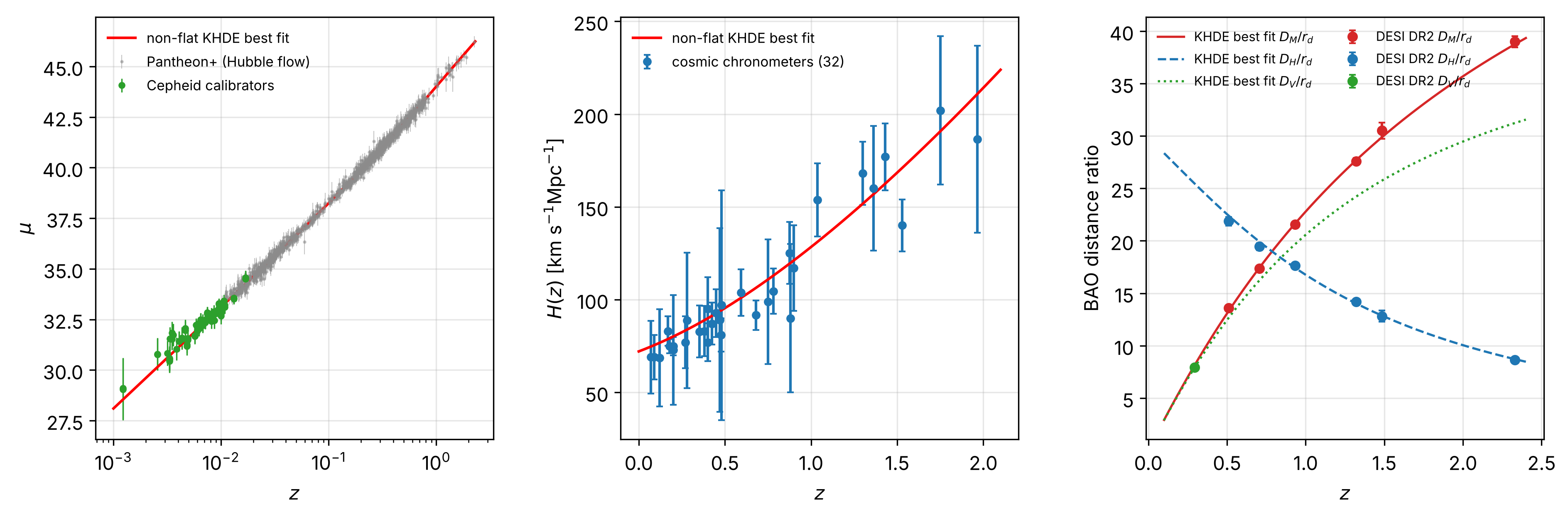}
\caption{The three datasets compared with the best-fit non-flat KHDE model
(red), evaluated at the joint maximum-likelihood parameters. Left:
Pantheon+SH0ES distance moduli at the best-fitting absolute magnitude, with the
Cepheid calibrators in green. Centre:
the $32$ cosmic-chronometer measurements of $H(z)$. Right: the DESI~DR2 BAO
ratios, with the model predictions for $D_M/r_d$ (solid), $D_H/r_d$ (dashed)
and $D_V/r_d$ (dotted).}
\label{fig:data}
\end{figure}

Figure~\ref{fig:data} shows each dataset alongside the best-fit non-flat KHDE
model, in the observable each probe actually measures: the supernova Hubble
diagram, the chronometer $H(z)$ points, and the DESI ratios $D_M/r_d$ and
$D_H/r_d$. The model reproduces all three without strain, which is the
expected counterpart of the $\chi^2/\mathrm{dof}=\rsChiRed$ reported below.

\subsection{Sampling}

The total likelihood is the product of the three, and we sample
$(c,K,\Om,\Ok,H_0,r_d)$ with the affine-invariant ensemble sampler
\texttt{emcee} \cite{Foreman-Mackey:2012any} under flat priors
$c\in[0.6,2.0]$, $K\in[0,0.8]$, $\Om\in[0.10,0.50]$, $\Ok\in[-0.30,0.30]$,
$H_0\in[60,85]$ and $r_d\in[120,180]\,\mathrm{Mpc}$, intersected with the
validity domain of Sec.~\ref{sec:numerics}\,(iv); the supernova absolute
magnitude is marginalised through Eq.~(\ref{eq:chisnmarg}). Convergence
diagnostics are reported in Sec.~\ref{sec:results}. Maximum-likelihood points,
which enter only the model comparison and Fig.~\ref{fig:data}, are located by
evaluating $\chi^2$ on $3000$ posterior samples and polishing the six lowest
with Nelder--Mead followed by bounded Powell minimisation; the reason is given
in Sec.~\ref{sec:validation}.

\section{Results}
\label{sec:results}

\subsection{Joint constraints}

\begin{figure}[H]
\centering
\includegraphics[width=0.92\linewidth]{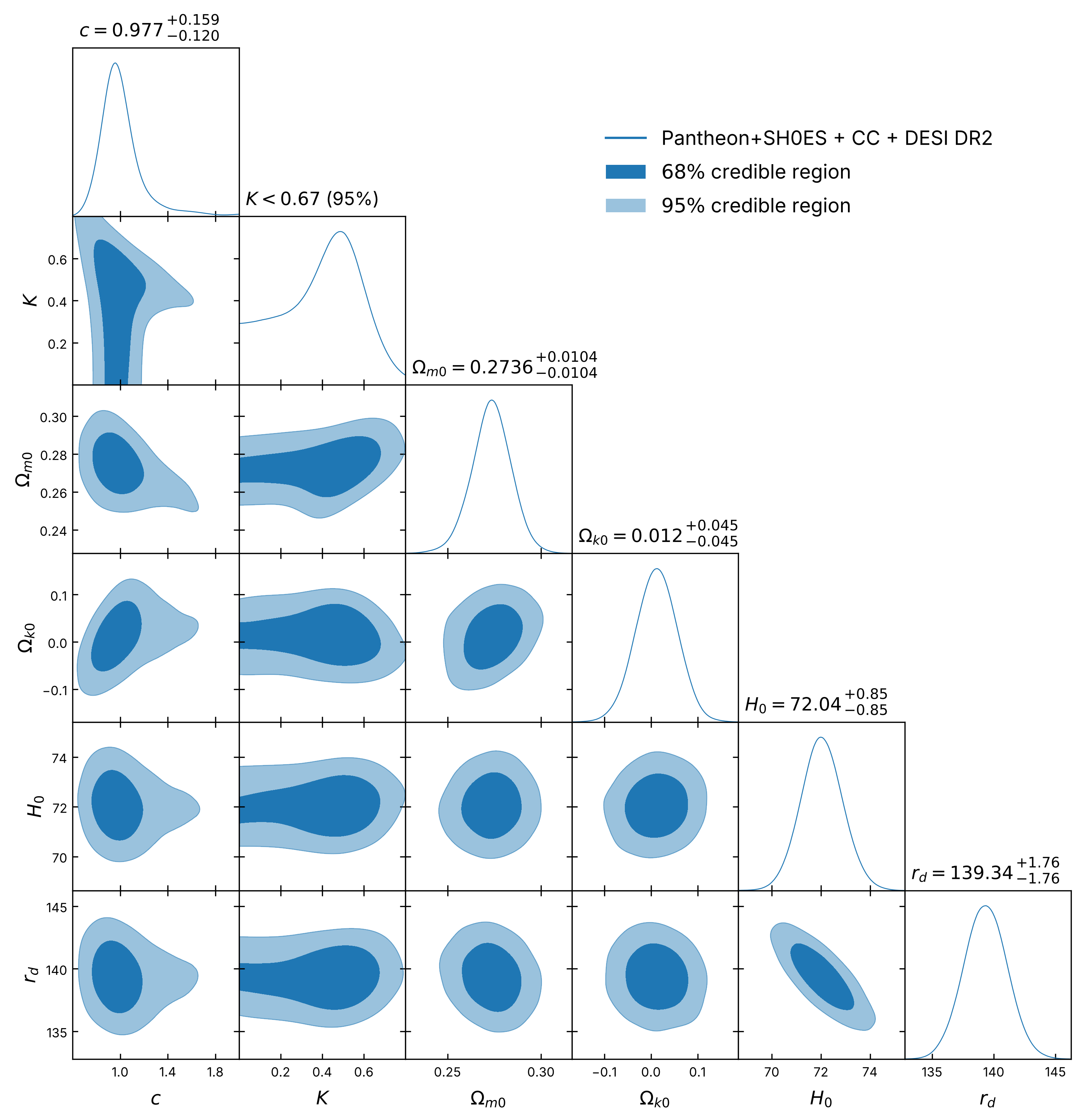}
\caption{Marginalised posterior for the non-flat KHDE model from the joint
Pantheon+SH0ES, cosmic-chronometer and DESI~DR2 dataset, showing $68\%$ and
$95\%$ credible regions. Panel headings give the medians with $68\%$ credible
intervals, as in Table~\ref{tab:joint}; for $K$ the one-sided $95\%$ upper
limit is shown.}
\label{fig:joint}
\end{figure}

\begin{figure}[H]
\centering
\includegraphics[width=0.62\linewidth]{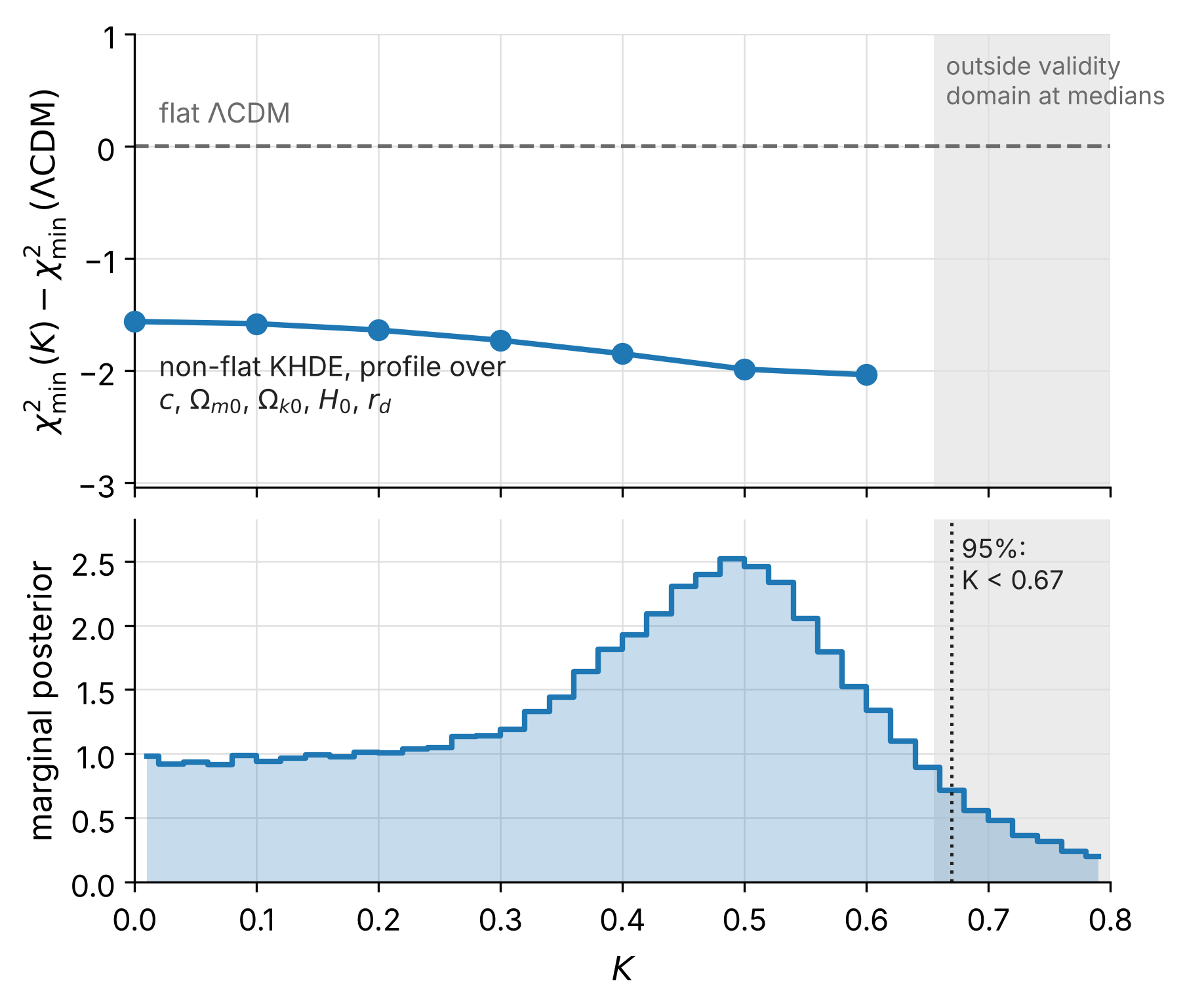}
\caption{Is the Kaniadakis deformation detected? Top: minimum $\chi^2$ of non-flat
KHDE at fixed $K$, minimised over $c$, $\Om$, $\Ok$, $H_0$ and $r_d$, relative to
the flat-$\Lambda$CDM minimum on the same joint data (dashed). Bottom: marginal
posterior of $K$, with its $95\%$ upper limit (dotted). Shaded: $K$ beyond the
validity cap $A^2\ge12c^2K^2M_p^4$ at the posterior medians. The data are fitted
equally well at every $K$; the shape of the posterior reflects how much of the
space of the other parameters fits at each $K$, and its upper limit the validity
domain.}
\label{fig:profile}
\end{figure}

Figure~\ref{fig:joint} shows the marginalised posterior and
Table~\ref{tab:joint} the resulting constraints. The joint dataset gives
\begin{equation}
\Om=\rsJOm,\quad
\Ok=\rsJOk,\quad
H_0=\rsJH,\quad
r_d=\rsJrd\ \mathrm{Mpc},
\end{equation}
with the holographic parameter $c=\rsJc$ and the Kaniadakis parameter
unconstrained, $K<\rsJK$ at $95\%$ confidence. The profile likelihood makes the
situation explicit. Minimising $\chi^2$ over the other five parameters at fixed
$K$, the minimum changes by only $\Delta\chi^2=\rsProfK$ between $K=0$ and the
best-fitting $K$ (Fig.~\ref{fig:profile}): the data are fitted equally well with
and without the deformation. The broad maximum of the marginal posterior near
$K\approx0.5$ is a projection effect rather than a preference of the data: there
a wider range of $c$ fits equally well, the conditional spread of $c$ being
nearly twice that at $K\simeq0$, so marginalising over $c$ gives that region
more weight. Above $K\simeq0.65$ the validity domain of
Sec.~\ref{sec:numerics}\,(iv) removes most of the allowed range of $c$, and it is
that cut, not the data, which sets the $95\%$ limit. We therefore quote no
two-sided interval.

The chain used $40$ walkers advanced for $1.8\times10^{4}$ steps with the first
$4500$ discarded, giving $5.4\times10^{5}$ samples at a mean acceptance fraction
of $\rsAcc$. The integrated autocorrelation times are $\tau=\rsTau$, so the
chain runs for $\rsSpt\,\tau$, in every case above the length of $50\,\tau$
usually required; the slowest-mixing parameter is $c$, whose degeneracy with
both $K$ and $\Om$ is visible in Fig.~\ref{fig:joint}. Splitting the walkers
into two halves gives Gelman--Rubin statistics $\hat R\le\rsRhat$ for every
parameter, against the customary threshold $\hat R<1.01$. The effective sample
size is $\rsEss$ per parameter, so the Monte Carlo error on each quoted median
is below $\rsMcse\,\sigma$, an order of magnitude smaller than the precision at
which the results are reported. The supernova-only and CC + DESI~DR2 chains of
Sec.~\ref{sec:complement}, with $1.4\times10^{4}$ steps each, run for at least
$\rsSptS\,\tau$ and $\rsSptB\,\tau$ respectively.

\begin{table}[H]
\centering
\caption{Constraints on the non-flat KHDE model. The supernova-only column
uses Pantheon+SH0ES alone; the joint column adds cosmic chronometers and
DESI~DR2. Values are medians with $68\%$ credible intervals; for $K$, which is
not detected, the one-sided $95\%$ upper limit is given.}
\label{tab:joint}
\begin{tabular}{lccc}
\toprule
Parameter & Prior & SN only & SN + CC + BAO\\
\midrule
$c$    & $\mathcal{U}(0.6,2.0)$   & $\rsSc$ & $\rsJc$\\
$K$    & $\mathcal{U}(0.0,0.8)$   & $<\rsSK$ ($95\%$)           & $<\rsJK$ ($95\%$)\\
$\Om$  & $\mathcal{U}(0.10,0.50)$ & $\rsSOm$            & $\rsJOm$\\
$\Ok$  & $\mathcal{U}(-0.30,0.30)$& $\rsSOk$ & $\rsJOk$\\
$H_0$ [km s$^{-1}$Mpc$^{-1}$] & $\mathcal{U}(60,85)$ & $\rsSH$ & $\rsJH$\\
$r_d$ [Mpc] & $\mathcal{U}(120,180)$ & --- & $\rsJrd$\\
\bottomrule
\end{tabular}
\end{table}

\subsection{What each probe contributes}
\label{sec:complement}

\begin{figure}[H]
\centering
\includegraphics[width=\linewidth]{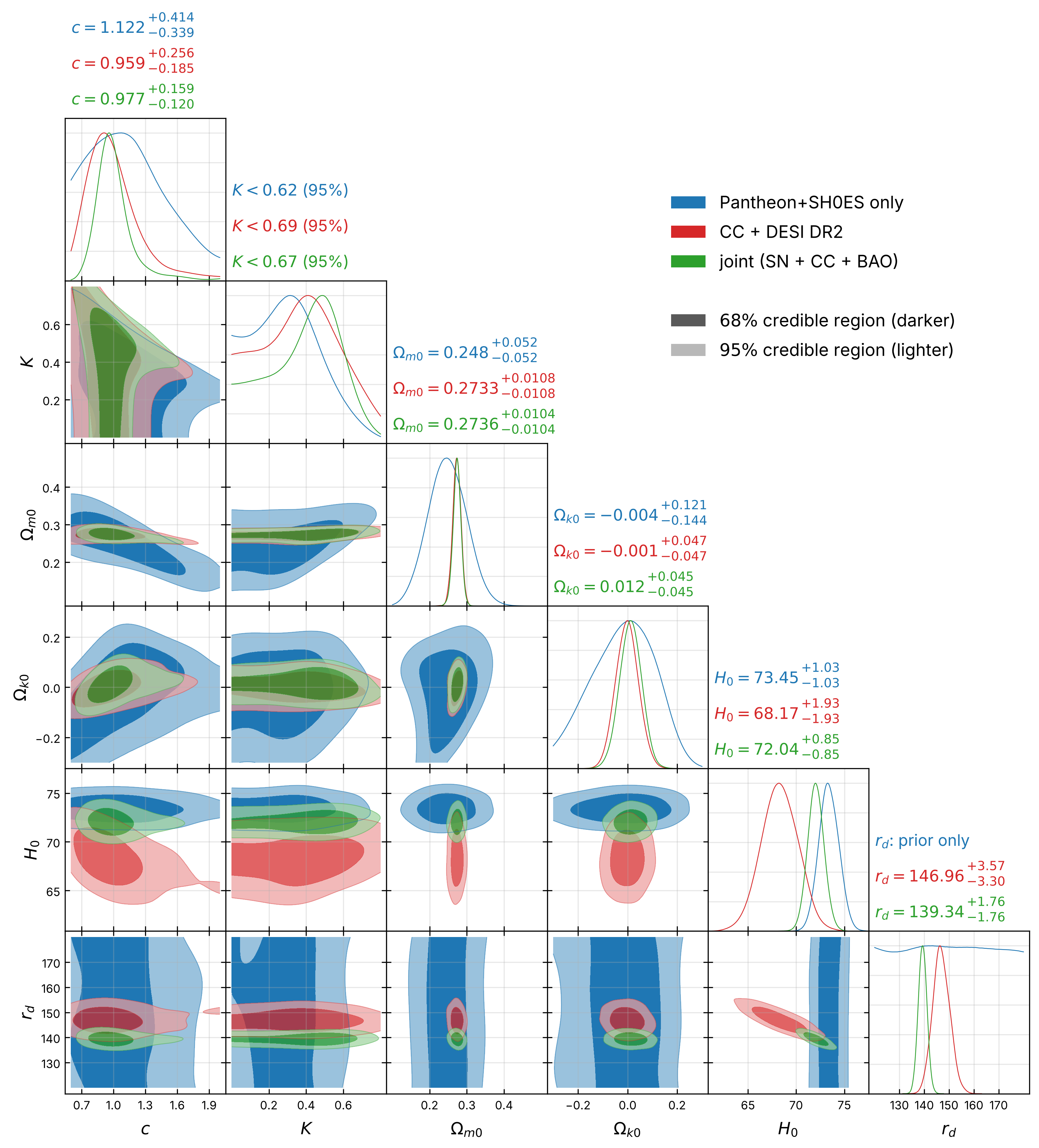}
\caption{Posterior distributions from Pantheon+SH0ES alone (blue), from
CC + DESI~DR2 (red), and from the joint dataset (green), showing $68\%$ and
$95\%$ credible regions. The values above each diagonal panel are those of
Table~\ref{tab:indiv}, in the colour of the corresponding dataset. The $r_d$
panel for supernovae alone reproduces the prior, and the
$H_0$--$r_d$ panel for CC + BAO shows the degeneracy that the supernova
calibration breaks.}
\label{fig:compare}
\end{figure}

Constraints from the individual probes are compared with the joint result in
Fig.~\ref{fig:compare} and Table~\ref{tab:indiv}. Each dataset fails in a
characteristic way, and the pattern of failures is what makes the combination
work.

\begin{table}[H]
\centering
\caption{Constraints from the probes separately and jointly (medians with
$68\%$ intervals; one-sided $95\%$ upper limits for $K$). Supernovae alone carry no information on $r_d$, so that entry
reproduces its prior.}
\label{tab:indiv}
\begin{tabular}{lccc}
\toprule
Parameter & Pantheon+SH0ES & CC + DESI DR2 & joint\\
\midrule
$c$    & $\rsSc$ & $\rsBc$ & $\rsJc$\\
$K$    & $<\rsSK$ ($95\%$) & $<\rsBK$ ($95\%$) & $<\rsJK$ ($95\%$)\\
$\Om$  & $\rsSOm$   & $\rsBOm$ & $\rsJOm$\\
$\Ok$  & $\rsSOk$ & $\rsBOk$ & $\rsJOk$\\
$H_0$ [km s$^{-1}$Mpc$^{-1}$] & $\rsSH$ & $\rsBH$ & $\rsJH$\\
$r_d$ [Mpc] & prior only & $\rsBrd$ & $\rsJrd$\\
\bottomrule
\end{tabular}
\end{table}

Calibrated supernovae determine $H_0=\rsSH$ but say nothing about $r_d$: their
posterior for it is flat across the prior, as it must be, since $r_d$ does not
enter the supernova likelihood at all. They also constrain curvature only to
$\Ok=\rsSOk$.

The two $H(z)$-sensitive probes do the opposite. CC + DESI~DR2 tighten curvature
by a factor of three, to $\Ok=\rsBOk$, and determine $r_d$ to a few per
cent, but leave $H_0$ loose at $\rsBH$ along a pronounced
$H_0$--$r_d$ degeneracy, visible in the $H_0$--$r_d$ panel (bottom row) of
Fig.~\ref{fig:compare}. This is the familiar statement that BAO measures the
product $H_0r_d$.

\subsection{The Hubble tension in this analysis}

The comparison makes an independent point that is worth stating explicitly.
Without the Cepheid calibration, CC + DESI~DR2 prefer
\begin{equation}
H_0=\rsBH\ \mathrm{km\,s^{-1}\,Mpc^{-1}},
\qquad r_d=\rsBrd\ \mathrm{Mpc},
\end{equation}
both remarkably close to the Planck $\Lambda$CDM values, $H_0\simeq67.4$ and
$r_d\simeq147.05\,\mathrm{Mpc}$ \cite{Planck:2018vyg}, even though no CMB
information enters our analysis. Calibrated supernovae alone give
$H_0=\rsSH$. The two determinations rest on independent data and differ by
$\rsHdiff\pm\rsHsig\,\mathrm{km\,s^{-1}\,Mpc^{-1}}$, a $\rsHtension\sigma$
discrepancy: the Hubble tension in miniature. The joint fit settles between
them, at $H_0=\rsJH$ with $r_d=\rsJrd\,\mathrm{Mpc}$, and the shift in $r_d$ is
the same tension expressed in the sound horizon rather than the expansion rate.

The KHDE freedom does not resolve it: allowing $c$, $K$ and $\Ok$ to vary leaves
the discrepancy in place, consistent with the general finding that late-time
modifications of the dark-energy sector struggle to reconcile the two
calibrations \cite{DiValentino:2021izs}.

\subsection{Curvature}

The improvement in the curvature constraint is the clearest benefit of the
combination. From supernovae alone we obtain
$\Ok=\rsSOk$, an interval of width $0.26$ that excludes
essentially nothing within the prior. Adding the two $H(z)$-sensitive probes
gives $\Ok=\rsJOk$, a factor of three tighter, and consistent with
spatial flatness at well under $1\sigma$. The reason is structural rather than
statistical: supernovae measure only the transverse distance, whereas BAO
supplies $D_H/r_d$ and $D_M/r_d$ separately and chronometers supply $H(z)$
without a distance ladder, and it is the radial-to-transverse ratio that
responds to curvature.

We note that the constraint remains an order of magnitude weaker than the
CMB+BAO bound $|\Omega_k|\lesssim0.002$. That is expected: the present
combination deliberately avoids CMB information, so as not to import a
$\Lambda$CDM-calibrated sound horizon into a test of non-standard dark energy.

\subsection{The sound horizon and \texorpdfstring{$H_0$}{H0}}

Keeping $r_d$ free yields $r_d=\rsJrd\,\mathrm{Mpc}$, some $\rsRdPlanck\%$ below the
Planck value $\simeq147\,\mathrm{Mpc}$. This is the familiar consequence of
anchoring the absolute scale to the Cepheid calibration: BAO constrains the
product $H_0r_d$: a flat-$\Lambda$CDM fit to DESI alone gives
$r_dh\simeq101.5\,\mathrm{Mpc}$, and the joint KHDE posterior gives
$r_dh=\rsRdh\,\mathrm{Mpc}$, so a high $H_0$ necessarily implies a low
$r_d$. The result should be
read as a statement about that degeneracy, not as an independent measurement of
the drag-epoch physics.

\subsection{Model comparison}

With the absolute magnitude marginalised in both models, flat $\Lambda$CDM
fitted to the same combination gives $\chi^2=\rsChiL$ for $\rsDofL$ degrees of
freedom, decomposed as $\rsChiLsn$ (SN) $+\,\rsChiLcc$ (CC) $+\,\rsChiLbao$
(BAO), against $\chi^2=\rsChiK$ for $\rsDofK$ for non-flat KHDE, decomposed as
$\rsChiKsn$ $+\,\rsChiKcc$ $+\,\rsChiKbao$. The non-flat model thus fits
marginally better, by $\Delta\chi^2=\rsDchiJ$, an improvement comparable to
what three additional parameters typically produce from noise alone. The
information criteria therefore favour $\Lambda$CDM,
\begin{equation}
\Delta\mathrm{AIC}=\rsDaicJ,\qquad \Delta\mathrm{BIC}=\rsDbicJ ,
\end{equation}
which on the scale conventionally used in cosmology \cite{Liddle:2007fy}, where
a difference of $2$ counts as positive and $6$ or more as strong evidence, is
positive evidence by AIC and strong evidence by BIC. Both verdicts come entirely
from the parameter count, not from a poorer fit.

Supernovae alone give almost the same comparison, $\Delta\chi^2=\rsDchiS$,
$\Delta\mathrm{AIC}=\rsDaicS$ and $\Delta\mathrm{BIC}=\rsDbicS$. Adding the
$H(z)$-sensitive probes tightens the curvature and separates $H_0$ from $r_d$,
but it does not discriminate between the two expansion histories: in every
combination the extra freedom of the model is permitted but not required.

\FloatBarrier
\section{Background evolution across the joint posterior}
\label{sec:background}

\begin{figure}[H]
\centering
\includegraphics[width=\linewidth]{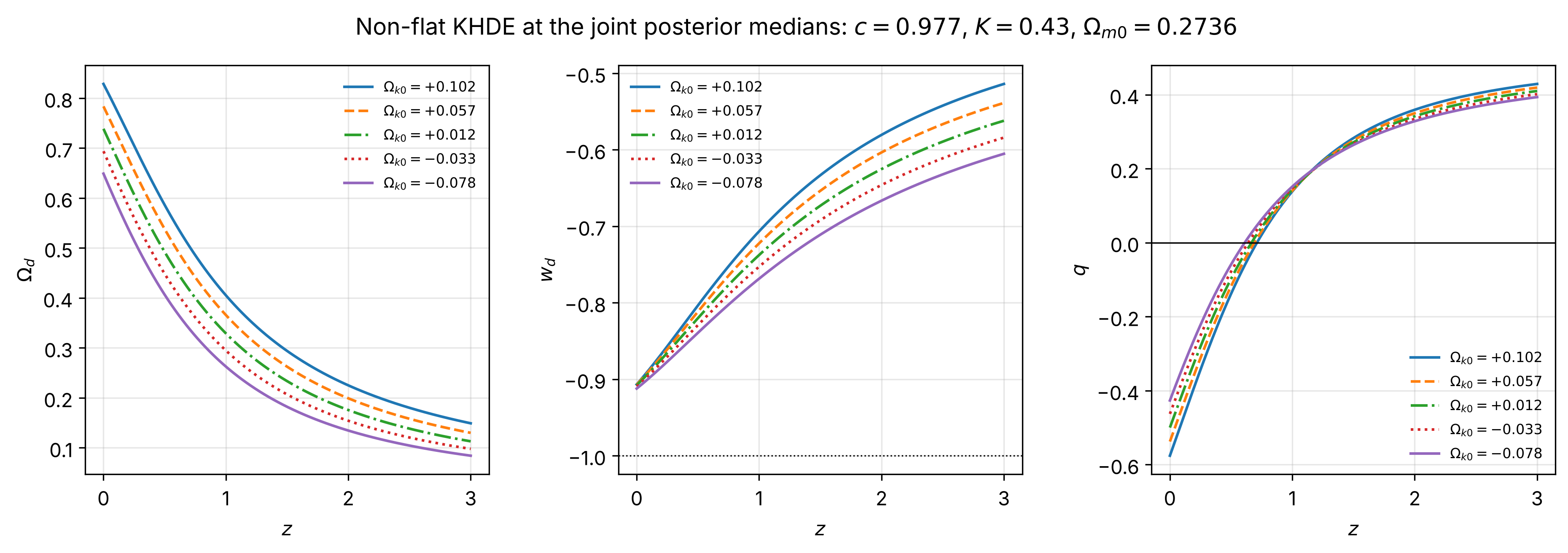}
\caption{Background evolution of the non-flat KHDE model at the joint
posterior medians $c=\rsMedc$, $K=\rsMedK$ and $\Om=\rsMedOm$, for
$\Ok=\rsMedOk\pm1\sigma,\pm2\sigma$. Left:
dark-energy density parameter. Centre: equation of state. Right: deceleration
parameter, whose sign change marks the onset of acceleration.}
\label{fig:bg}
\end{figure}

Having fixed the parameters, we examine what the model predicts.
Figure~\ref{fig:bg} shows the dark-energy density parameter, the equation of
state and the deceleration parameter at the joint posterior medians of $c$, $K$
and $\Om$, with the curvature set to its median and to $\pm1\sigma$ and
$\pm2\sigma$ about it. Because the likelihood is flat in $K$, the medians are a
representative choice for displaying the deformed model; derived quantities
quoted below are propagated from $2500$ posterior samples rather than read off
the curves.

Three features deserve comment.

\emph{The equation of state lies in the quintessence regime.} Propagating the
posterior gives $\wde(z=0)=\rsWz$, and $\wde$ increases monotonically
towards higher redshift for every sample. For $\rsNonPhantom\%$ of the posterior
the model stays above $-1$ throughout $0\le z\le3$; the remaining
$\rsPhantom\%$, all with $c<\rsPhantomC$, dip below $-1$ only at the lowest
redshifts, where $\wde$ attains its
minimum, as expected
for holographic dark energy with $c<1$ and a future-event-horizon cutoff
\cite{Li:2004rb}. The data therefore favour, but do not require, a
non-phantom equation of state.

\emph{Curvature acts oppositely on $\wde$ and on $q$.} Increasing $\Ok$ towards
the closed branch raises $\Od$ today, from $\rsOdOpen$ to $\rsOdClosed$ across
the plotted range. This leaves $\wde(0)\simeq\rsWzMed$ almost unchanged but
drives $\wde$ upward at $z\gtrsim0.5$, while it drives $q$ downward at low
redshift, from $q_0=\rsQzOpen$ to $\rsQzClosed$, because the explicit curvature
term in
Eq.~(\ref{eq:q}) enters with the same sign as the dark-energy contribution. The
two effects partly cancel in the distance observables, which is one reason
supernovae alone constrain $\Ok$ so weakly.

\emph{The transition redshift is stable.} Across the posterior the
deceleration-to-acceleration transition occurs at $z_t=\rsZt$
($\rsZtLo$--$\rsZtHi$ at $95\%$), and the present deceleration parameter is
$q_0=\rsQz$; the transition is consistent with the flat-$\Lambda$CDM
value $z_t\simeq0.67$ for $\Omega_m\simeq0.30$. The model is
thus not distinguished from the concordance picture by its expansion
kinematics, which is consistent with the model comparison of
Sec.~\ref{sec:results}: what counts against it is the extra freedom without a
corresponding improvement in fit.

\subsection{Sensitivity to the Kaniadakis parameter}

Why the data leave $K$ unconstrained can be seen in two steps. First, even with
$c$, $\Om$, $\Ok$, $H_0$ and $r_d$ held at their posterior medians, the
deformation barely changes the observables: the $\chi^2$ distance between the
predictions at $K$ and at $K=0$, under the joint covariance and with the
absolute magnitude marginalised, is $\rsSensA$, $\rsSensB$, $\rsSensC$ and
$\rsSensE$ for $K=0.1,0.2,0.3$ and $0.5$. The Kaniadakis term is comparable to
$A^2$ only at low redshift, where $A$ is smallest, and only values near the cap
set by the validity condition of Sec.~\ref{sec:numerics}\,(iv) produce a
difference of order unity. Second, that residual signal is absorbed:
re-minimising $\chi^2$ over the other five parameters at each fixed $K$, the
joint $\chi^2$ changes by only $\rsProfK$ between $K=0$ and the best-fitting $K$
(Fig.~\ref{fig:profile}). The deformation is both small and degenerate with
shifts in $c$, $\Om$ and $\Ok$, and current data cannot separate them.

\FloatBarrier
\FloatBarrier
\section{Model-independent reconstruction}
\label{sec:ann}

The constraints above assume the KHDE form of the expansion history. As a
complementary check we now reconstruct $H(z)$ directly from data with an
artificial neural network, imposing no cosmological model, and ask whether the
fitted model is consistent with it. Neural networks have also been applied to
KHDE to estimate its parameters by likelihood-free inference
\cite{Kapil:2025mpt,Kapil:2025pyt,Kapil:2026hcd,Singh:2026qfj}; here the network
plays a different role, since it is trained on the data alone and never sees the
model.

\subsection{Why the reconstruction must use chronometers}

In a flat universe one can reconstruct $H(z)$ from supernovae alone: the
comoving distance is $D_C=D_L/(1+z)$ and $H=c_{\rm light}/(\mathrm{d}D_C/\mathrm{d}z)$,
with no further input. That inversion fails once curvature is allowed. With
$\Ok\ne0$ the measured quantity is $D_M=(c_{\rm light}/H_0)S_k(\chi)$, so
recovering the radial coordinate requires
$\chi=S_k^{-1}(D_MH_0/c_{\rm light})$ and hence a value of $\Ok$. A
supernova-based reconstruction of $H(z)$ in a curved universe is therefore
conditional on the curvature rather than independent of it, and cannot be used
to test a model in which $\Ok$ is a free parameter.

Cosmic chronometers have no such difficulty. Equation~(\ref{eq:cc}) returns
$H(z)$ from the differential-age relation with no integral, no distance ladder
and no assumption about curvature or the dark-energy sector. The curvature-free
reconstruction in a non-flat analysis is thus the one trained on chronometer
data, and that is what we use.

\subsection{Network and training}

We use a bootstrap deep ensemble of $40$ networks mapping
$z\mapsto[H(z),\ln\sigma_{\rm int}(z)]$, each trained on an independent
resample of the $32$ measurements under the heteroscedastic Gaussian negative
log-likelihood. With so few data points the architecture must be small and
strongly regularised: a single hidden layer of eight $\tanh$ units with weight
decay $10^{-1}$. Larger networks reproduce the data equally well but develop
spurious structure between points, which is harmless for $H(z)$ itself and
fatal for its derivative; in testing, an unregularised network with two layers
of $32$ units gave a root-mean-square error in $q(z)$ five times larger, with an
apparent acceleration--deceleration transition displaced to $z\simeq1.2$. The
derivative $\mathrm{d}H/\mathrm{d}z$ is obtained by automatic differentiation of
each ensemble member, and the deceleration parameter follows from the
model-independent relation
\begin{equation}
q(z)=(1+z)\frac{H'(z)}{H(z)}-1 .
\label{eq:qrec}
\end{equation}
We evaluate Eq.~(\ref{eq:qrec}) separately for each ensemble member and quote
the median and standard deviation across the ensemble. This retains the
correlation between $H$ and $H'$ within each network; propagating the ensemble
errors of $H$ and $H'$ independently would misstate the uncertainty, and in
particular would make it vanish wherever $q$ crosses zero.

\subsection{Results}

\begin{figure}[H]
\centering
\includegraphics[width=\linewidth]{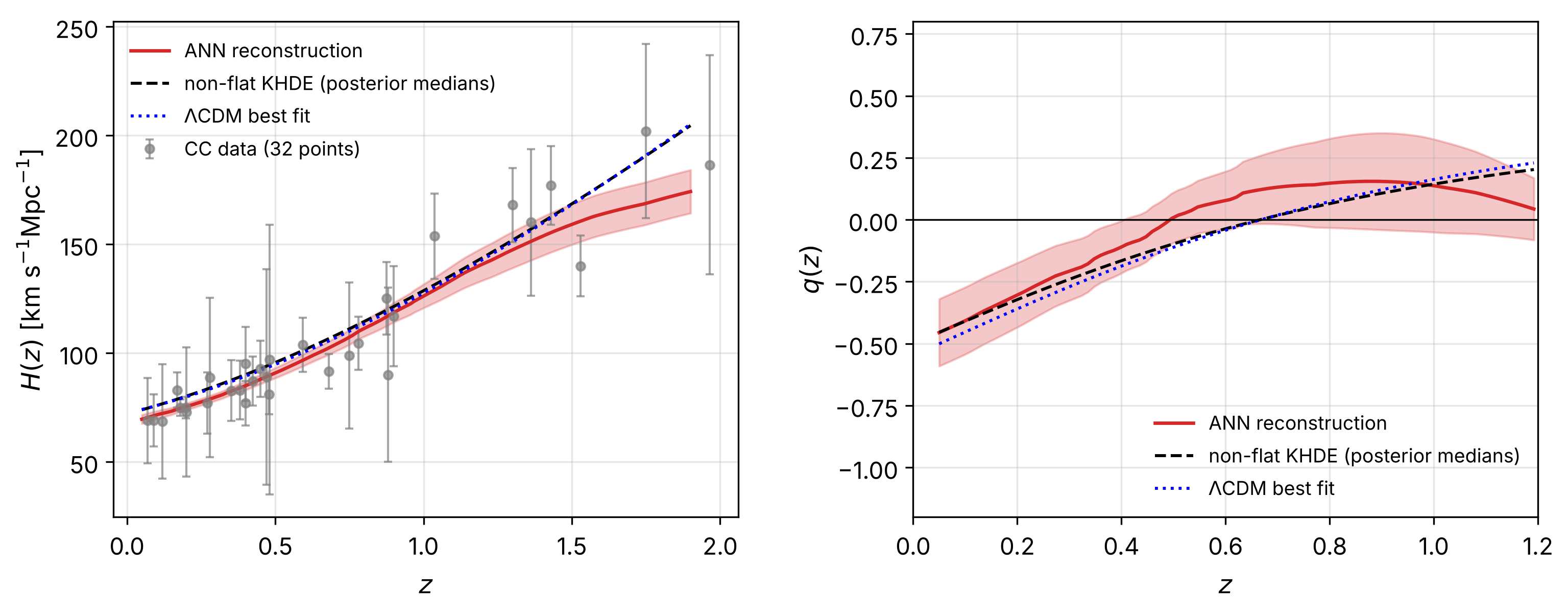}
\caption{Model-independent ANN reconstruction from the $32$ cosmic-chronometer
measurements, with the $1\sigma$ ensemble band. Left: the expansion rate, with
the data, the non-flat KHDE model at the joint posterior medians (dashed) and
the flat-$\Lambda$CDM best fit (dotted). Right: the deceleration parameter from
Eq.~(\ref{eq:qrec}), shown over $z<1.2$ where the derivative is supported by the
data.}
\label{fig:ann}
\end{figure}

Figure~\ref{fig:ann} shows the reconstruction. At the lowest point of the
reconstruction grid, $z=0.05$, the ensemble gives
$H=\rsAnnH\,\mathrm{km\,s^{-1}\,Mpc^{-1}}$,
consistent with the CC + BAO value of Sec.~\ref{sec:complement} and lying
between the two calibrations discussed there. The reconstructed expansion rate
agrees with the non-flat KHDE model evaluated at the joint posterior medians
to $\rsAnnK\%$ on average over $0.1<z<1.5$, and with the flat-$\Lambda$CDM best
fit to $\rsAnnL\%$; the two models are not distinguished by the chronometer
data, consistent with the model comparison of Sec.~\ref{sec:results}, in which
neither fits better than the other.

The reconstructed deceleration parameter at the lowest grid point, $z=0.05$,
is $q=\rsAnnQ$, which we quote as $q_0$; the offset from $z=0$ is small
compared with the uncertainty. All $40$ networks give $q_0<0$, so present
acceleration is recovered at $\rsAnnSig\sigma$ without assuming any model, and
the value agrees within $\rsAnnAgree\sigma$ with the KHDE posterior value
$q_0=\rsQz$ of Sec.~\ref{sec:background}. Both model curves lie within
the $1\sigma$ band over the plotted range; the reconstruction runs slightly
above them for $0.4\lesssim z\lesssim0.8$, and the band widens beyond
$z\simeq0.6$ as the data thin out. In $H(z)$ the reconstruction falls below
both models above $z\simeq1.4$, where only four chronometer points, including
the low value at $z=1.53$, constrain it; since the two models coincide there,
this reflects the edge of the data rather than a preference between them. We do not quote a reconstructed
transition redshift: with $32$ points the zero-crossing of $q$ is sensitive to
the regularisation at the level of $\Delta z\sim0.5$, and any value obtained
from it would say more about the network than about the Universe. This is a
limitation of the data rather than of the method, and it is the same limitation
that makes the chronometer contribution to the joint fit modest.

\section{Discussion}
\label{sec:discussion}

\subsection{What the model comparison does and does not show}

Two conclusions are easily conflated, and the analysis supports only one of
them. The first is that the data do not require the Kaniadakis deformation, nor
the additional freedom of curvature and of the holographic parameter: both
information criteria favour $\Lambda$CDM, and the profile likelihood is flat in
$K$. The second, which the data do not support, is that the model is excluded.
Its best fit is as good as that of $\Lambda$CDM on supernovae alone and on the
joint dataset, and its posterior expansion history is close to the concordance
one (Sec.~\ref{sec:background}). Non-flat KHDE is, on current data, an
observationally indistinguishable and therefore unnecessary extension, not a
falsified one.

The distinction matters for how such models are tested. An information-criterion
penalty for unconstrained parameters is a statement about economy, not about
consistency: a model can lose clearly on $\Delta\mathrm{BIC}$ while describing
every dataset to within its errors. Conversely, a large $\Delta\chi^2$ against a
model should be checked against the posterior samples before it is reported as
an exclusion. In this model the validity domain, the near-perfect $H_0$--$r_d$
degeneracy and the flat direction in $K$ make local optimisers unreliable, and
the failure is not benign: it produces precisely the kind of apparent rejection
that would be interpreted as evidence against the model
(Sec.~\ref{sec:validation}).

\subsection{What would change the conclusion}

Three things could alter the picture, and it is worth being explicit about them.

First, the chronometer covariance. We treat the $32$ measurements as
uncorrelated, whereas the stellar-population modelling systematics are shared
between subsets from the same analysis \cite{Moresco:2020fbm}. A full covariance
would inflate the CC uncertainties, loosening the curvature constraint and the
CC + DESI~DR2 value of $H_0$; it is unlikely to change the model comparison,
which is set by the parameter count rather than by the fit.

Second, the treatment of $r_d$. Imposing a Planck prior would tighten $H_0$
substantially, but at the cost of conditioning a test of non-standard dark
energy on $\Lambda$CDM early-Universe physics. Our choice is deliberately
conservative; a reader who accepts the Planck calibration should expect
correspondingly tighter, and differently centred, constraints.

Third, the supernova compilation. Pantheon+SH0ES is the only one of the
currently available samples carrying a Cepheid calibration, which is why we use
it, but Union3 \cite{Rubin:2023ovl} and DES~Y5 \cite{DES:2024tys} differ in
their calibration and redshift coverage, and model-independent analyses have
found the inferred dark-energy dynamics to be sensitive to that choice
\cite{Maqsood:2026thermo}. Repeating the analysis with each would test whether
the conclusions are robust to the supernova sample, and we regard this as the
most useful immediate extension.

\subsection{Relation to earlier work}

Non-flat KHDE was introduced with the apparent horizon as cutoff in
Ref.~\cite{Sharma:2021zjx}, where it was compared with chronometer and Union~2.1
data at fixed values of $\Ok$, and with the future event horizon in
Ref.~\cite{P:2022amn}, which analysed its background dynamics, treating the
closed and open branches separately and exploring the parameter dependence
without a joint likelihood analysis. The present work takes the event-horizon
equations and confronts them with current data, keeping $\Ok$ free rather than
selecting a branch by hand; the sign of $\Ok$ is then an output rather than an
input, and we find it consistent with zero. The flat version of the model has
been constrained with supernovae and chronometers previously
\cite{Drepanou:2021jiv,Hernandez-Almada:2021aiw,Sharma:2023til}, with results
for $c$ and $\Om$ broadly consistent with ours once the different data
combinations are allowed for. The recent flat analyses with DESI~DR2 reach a
similar verdict: they report no significant evidence for the deformation
\cite{Luciano:2025ykr,Li:2025ivi}, and information criteria or the Bayesian
evidence prefer $\Lambda$CDM once the extra parameters are accounted for
\cite{Luciano:2025ykr,Gonzalez-Espinoza:2026iyh}. Our result shows that freeing
the curvature and calibrating the supernovae on the Cepheids do not change that
verdict.

Standard HDE, the $K\to0$ limit of the model, has been reported as strongly
disfavoured relative to $\Lambda$CDM in a flat universe, with DESI~DR1 and
either Union3 supernovae with an SH0ES prior on $H_0$ or the Cepheid-calibrated
Pantheon+SH0ES sample \cite{Zapata:2025ngr}. In our analysis the $K=0$ model,
with curvature free, fits slightly better than flat $\Lambda$CDM, with
$\Delta\chi^2=\rsProfKzL$ (Fig.~\ref{fig:profile}). The two analyses differ in curvature, in the
chronometers, in the DESI release and in the sound horizon:
Ref.~\cite{Zapata:2025ngr} imposes a Planck and BBN constraint on
$\Omega_bh^2$, which in a late-time analysis enters only through $r_d$, whereas
we leave $r_d$ free. We have not isolated which of these differences is
responsible.

\FloatBarrier
\section{Numerical validation of the implementation}
\label{sec:validation}

Because the non-flat KHDE equations are stiff in places and contain several
opportunities for silent error, we record here the checks performed on the
implementation and on the inference.

\subsection*{Analytic limits}

Setting $K\to0$ and $\Ok=0$ in Eqs.~(\ref{eq:ode}) and (\ref{eq:w}) must return
the standard flat HDE expressions \cite{Li:2004rb,Wang:2016och},
\begin{equation}
\Od'=\Od(1-\Od)\left(1+\frac{2}{c}\sqrt{\Od}\right),
\qquad
\wde=-\frac{1}{3}\left(1+\frac{2}{c}\sqrt{\Od}\right).
\label{eq:hdelimit}
\end{equation}
Evaluating both sides at $\Od=0.3,0.5,0.7,0.9$ with $c=1.15$ gives agreement to
$1.4\times10^{-15}$ for the equation of state and $8.3\times10^{-16}$ for the
density derivative, i.e.\ to machine precision.

\subsection*{Reduction to the flat model}

Setting $\Ok=0$ but retaining $K=0.30$ must reproduce the flat KHDE model. We
compared the non-flat integrator against an independently written flat-KHDE
integrator over $0\le z\le3$; evaluated on a dense grid over that range the two agree to
$8\times10^{-15}$ in $\Od(z)$ and $1\times10^{-15}$ in $\wde(z)$. Since the two codes were written from
different equations, this is a genuine cross-check rather than a self-consistency
test.

\subsection*{Energy conservation and the event horizon}

Two further tests do not rely on any published expression. First, the equation
of state must satisfy the dark-sector conservation law,
$\wde=-1-\tfrac13\,\mathrm{d}\ln(\Od E^2)/\mathrm{d}x$, and the deceleration
parameter must satisfy $q=-1-\mathrm{d}\ln E/\mathrm{d}x$. Second, the cutoff
$L=\sqrt{6c^2/(A+B)}$ must obey the event-horizon relation
$\mathrm{d}L/\mathrm{d}x=L-\mathcal{C}/E$, which follows directly from
$L=a\,r(t)$ with $r=\sin y/\sqrt{k}$. With Eq.~(\ref{eq:w}) as given, all three
hold to better than $2\times10^{-8}$ for $\Ok=0,\pm0.05,\pm0.1$, with and without
the Kaniadakis term, confirming that Eqs.~(\ref{eq:ode}), (\ref{eq:H}),
(\ref{eq:w}) and (\ref{eq:q}) are mutually consistent.

\subsection*{Locating the maximum likelihood}

The model comparison rests on $\chi^2_{\min}$, so the minimum must be found
reliably, and in this model a local optimiser is not reliable. Outside the
validity domain $A^2\ge12c^2K^2M_p^4$ there is no likelihood at all, $H_0$ and
$r_d$ are almost perfectly degenerate in the BAO term, and the likelihood is flat
in $K$. Started from six generic points, Powell's method stopped
$\rsOptGap$ units of $\chi^2$ above the minimum, which would have turned a model
that fits as well as $\Lambda$CDM into one that appears to fit significantly
worse. We therefore evaluate $\chi^2$ on $3000$ posterior samples and
polish the six lowest with Nelder--Mead followed by bounded Powell minimisation.
By construction the result never lies above the best sample, and repeating the
procedure with an independent subsample reproduces $\chi^2_{\min}$ to
$0.01$.

\subsection*{Supernova absolute magnitude}

Using the released distance moduli as data fixes the absolute magnitude at
$M=-19.253$, so the $77$ calibrator residuals become constants and the
calibration uncertainty never reaches $H_0$. Flat $\Lambda$CDM then gives
$H_0=73.23\pm0.23$; with $M$ marginalised through Eq.~(\ref{eq:chisnmarg}) it
gives $H_0=\rsValH$, whose uncertainty matches the published one
(Table~\ref{tab:valid}). The joint fit places the absolute magnitude at
$M=\rsMB$.

\subsection*{Likelihood validation}

Each likelihood was validated against its published flat-$\Lambda$CDM result
before combination, as reported in Table~\ref{tab:valid}. In addition, the BAO
covariance was checked for symmetry and positive-definiteness, and the
correlation structure was verified to reproduce the anticorrelated $D_M$--$D_H$
pairs of the DESI release, with $\rho$ between $-0.35$ and $-0.49$.

\subsection*{Integrator tolerance}

The background ODE is integrated with an explicit Runge--Kutta scheme. Reducing
the tolerance from $\mathrm{rtol}=10^{-8}$ to $10^{-6}$, which speeds the
likelihood by roughly a factor of two, changes the predicted distance modulus by
at most $1.1\times10^{-5}$~mag across $0\le z\le3$, more than two orders of
magnitude below the typical supernova uncertainty divided by $\sqrt{N}$. The looser tolerance was
therefore used in the sampling, and the maximum-likelihood fit was verified to
reproduce the tighter-tolerance result to $0.01$ in $\chi^2$.

\subsection*{Convergence}

The joint chain satisfies both standard criteria: chain lengths of
$\rsSpt\,\tau$ against the recommended $50\,\tau$, and Gelman--Rubin statistics
$\hat R\le\rsRhat$ against the customary $\hat R<1.01$, with an effective sample
size of $\rsEss$ per parameter. The supernova-only and CC + DESI~DR2 chains run
for at least $\rsSptS\,\tau$ and $\rsSptB\,\tau$.

\section{Conclusions}
\label{sec:conclusions}

We have asked whether the Kaniadakis deformation of holographic dark energy is
detectable with current data in a spatially curved universe, using
Pantheon+SH0ES supernovae, $32$ cosmic chronometers and DESI~DR2 baryon acoustic
oscillations, with the sound horizon free so that no CMB calibration enters and
each likelihood validated against its published flat-$\Lambda$CDM result.

The answer is no. Non-flat KHDE fits the joint data as well as flat
$\Lambda$CDM, $\Delta\chi^2=\rsDchiJ$ for three additional parameters, but the
data leave the Kaniadakis parameter unconstrained: the profile $\chi^2$ changes
by only $\rsProfK$ between $K=0$ and its best value, the effect of $K$ on the
observables being absorbed by $c$, $\Om$ and $\Ok$, and the $95\%$ bound
$K<\rsJK$ is set by the validity domain of the model. Information criteria
therefore favour $\Lambda$CDM, $\Delta\mathrm{AIC}=\rsDaicJ$ and
$\Delta\mathrm{BIC}=\rsDbicJ$, entirely through the parameter count; supernovae
alone give the same picture. The model is permitted by the data but not
required.

The joint posterior gives $\Om=\rsJOm$, $\Ok=\rsJOk$,
$H_0=\rsJH\,\mathrm{km\,s^{-1}\,Mpc^{-1}}$, $r_d=\rsJrd\,\mathrm{Mpc}$ and
$c=\rsJc$, from a chain run for $\rsSpt\,\tau$ with $\hat R\le\rsRhat$. The
curvature constraint is three times tighter than from supernovae alone and
consistent with flatness. The derived expansion history,
$\wde(z=0)=\rsWz$, $q_0=\rsQz$ and $z_t=\rsZt$, is close to the concordance
one. Without the Cepheid calibration, chronometers and DESI prefer
$H_0=\rsBH$ and $r_d=\rsBrd\,\mathrm{Mpc}$, close to the Planck values and
$\rsHtension\sigma$ below the calibrated supernovae; the extra freedom of the
model does not reconcile the two. A neural-network reconstruction from the
chronometers, which in a curved universe cannot be built on supernovae because
inverting $D_M$ requires $\Ok$, gives $q_0=\rsAnnQ$, consistent with the model.

Three requirements of the analysis deserve emphasis because each changes the
result. The background equations must be solved in dimensionless form, without
which the deformation is numerically invisible. With the Cepheid-calibrated
supernova sample the absolute magnitude must be fitted, not fixed at the value
built into the released distance moduli, without which the uncertainty on $H_0$
is underestimated fourfold. And the maximum likelihood must be located from the
posterior samples, without which a local optimiser can report a statistically
equivalent model as excluded. We have also recorded the conventions on which
the implementation depends.

Detecting the deformation will require breaking its degeneracy with $c$, $\Om$
and $\Ok$. CMB distance priors would tighten the curvature by an order of
magnitude, at the cost of importing $\Lambda$CDM-calibrated information, and
repeating the analysis with the Union3 and DES~Y5 supernova samples would test
the robustness of these conclusions to the supernova calibration.

\section*{Acknowledgements}
The authors gratefully acknowledge the management of the University of
Technology and Applied Sciences, Salalah for providing the necessary
infrastructure and support.

\section*{Data availability}
This work uses publicly available datasets: the Pantheon+SH0ES compilation
\cite{Scolnic:2021amr,Brout:2022vxf}, the DESI~DR2 BAO measurements
\cite{DESI:2025zgx}, and the cosmic-chronometer compilation cited in
Sec.~\ref{sec:data}.

\section*{Conflict of interest}
The authors declare no competing interests.

\bibliographystyle{unsrt}
\bibliography{refPSK}

\end{document}